\documentclass[aps,prr,showpacs,preprintnumbers,twocolumn,superscriptaddress,longbibliography]{revtex4-1}
\usepackage{amsmath,amssymb}
\usepackage{bm}
\usepackage{tipa}
\usepackage{upgreek}
\usepackage{comment}
\usepackage{mathrsfs}
\usepackage{graphicx}
\usepackage{braket}
\usepackage{enumitem}
\usepackage{mathbbol}
\usepackage{gensymb}
\usepackage[normalem]{ulem}
\usepackage{color}
\usepackage[colorlinks,bookmarks=true,citecolor=blue,linkcolor=red,urlcolor=blue]{hyperref}
\usepackage{hyperref}
\usepackage{subfigure} 

\begin{document}

\title{Fractional diffusion without disorder in two dimensions}

\begin{abstract}
We analyse how simple local constraints 
in two dimensions lead a defect to exhibit robust,  non-transient, and tunable, subdiffusion. We uncover a rich dynamical phenomenology realised in ice- and dimer-type models. On the microscopic scale, a single defect exhibits anomalously long retractions. Such retractions result in a form of dynamical caging and can be  captured through an effective fractional diffusion equation. Mapping to a height field yields  an effective random walk subject to an emergent (entropic) logarithmic potential, whose strength is tunable, related to the exponent of algebraic ground-state correlations. The defect's path, viewed as non-equilibrium growth process, yields a frontier of fractal dimension of $5/4$, the value for a loop-erased random walk, rather than $4/3$ for simple and self-avoiding random walks. Such frustration/constraint-induced subdiffusion is expected to be relevant to platforms such as artificial spin ice and quantum simulators aiming to realize discrete link models and emergent gauge theories.
\end{abstract}

\author{Nilotpal Chakraborty}
\thanks{Corresponding author}
\email{nc553@cam.ac.uk}
\affiliation{Max-Planck-Institut f\"{u}r Physik komplexer Systeme, N\"{o}thnitzer Stra\ss e 38, Dresden 01187, Germany}
\affiliation{TCM Group, Cavendish Laboratory, University of Cambridge, Cambridge CB3 0HE, United Kingdom}

\author{Markus Heyl}
\affiliation{Theoretical Physics III, Center for Electronic Correlations and Magnetism, Institute of Physics, University of Augsburg, D-86135 Augsburg, Germany}

\author{Roderich Moessner}
\affiliation{Max-Planck-Institut f\"{u}r Physik komplexer Systeme, N\"{o}thnitzer Stra\ss e 38, Dresden 01187, Germany}

\maketitle

\section{Introduction}
Diffusion is a ubiquitous and universal phenomenon: the theory of hydrodynamics posits that conserved quantities generically diffuse \cite{chaikin1995principles}.
Deviations from this behaviour can nonetheless occur, for example as a result of disorder \cite{GB_subdiff}, fine-tuning of a system to the vicinity of an integrable \cite{Genhydro1,Genhydro2} or critical point \cite{anomdifffract,bunde2012fractals,hallen2022dynamical}, or for more complex conservation laws \cite{Fhydro1,FHydro2}. Origin, nature and robustness of the resulting sub-, superdiffusive, or indeed ballistic, transport are therefore natural objects of study.  Such anomalous diffusion processes have found several applications in fields ranging from biology to finance \cite{metzler2014anomalous}.

Conserved quantities may be of fundamental origin -- such as charge and 
mass of an elementary particle -- or emergent properties: simple lattice models of, say, Ising spins may yield cooperative behaviour the low-energy description of which includes an emergent conserved charge. Such models turn out to be quite commonly encountered in the study of frustrated and link models, whose low-energy description takes the form of an (emergent) U(1) gauge theory. This includes familiar representatives such as the ice and hardcore dimer models \cite{pauling1935structure,Liebice}. 

Recent interest in, and progress towards, realising gauge theories in quantum simulator platforms \cite{schweizer2019floquet,zhou2022thermalization,semeghini2021probing,meth2023simulating,martinez2016real,cochran2024visualizing,yang2020observation,gonzalez2024observation} calls for a study of the dynamics of such emergent charges \cite{DFL1,DFL2}. The basic question is: how do the constraints encoded by the gauge theory manifest themselves in the hydrodynamic behaviour of its conserved charges? In this work we consider perhaps the simplest such model, a link model with ice or dimer constraints \cite{Liebice,Fisherdimer,temperley1961dimer,kasteleyn1961statistics,baxter2007exactly}. We find that the motion of defects here 
presents an alternative to the conventional diffusive hydrodynamics. 

Concretely, we find a rich, {\it and tunable,} phenomenology of long-time subdiffusion which results from the constraints imposed by the emergent gauge fields. At the microscopic level, we identify a dynamical caging effect which makes a connection to loop-erased random walks and fractional diffusion. On the level of long-wavelength descriptions, we identify a continuum theory in the form of height models and a mapping to diffusion in a logarithmic potential, a topic of recent interdisciplinary interest \cite{angstmann2019time,Braylog,Lutzlog,campa2009statistical,bar2008dynamics,ray2020diffusion,onofri2020exploring}. The tunable anomalous  diffusion, arises purely from the residual entropy of the ground state manifold. Such dynamical phenomenology in the absence of any disorder, integrability or fine-tuning to a critical point, in simple and paradigmatic models of geometric frustration is a key result of our work.

Further, we find that the infinite-time subdiffusion for a single dynamical charge, is modified in the presence of a finite density 
of such mobile charges. In this case a length scale, corresponding to the average inter-particle spacing, is introduced upto which subdiffusion persists and beyond which there is crossover to ordinary diffusion. Finally, we examine the problem through the more mathematically motivated lens of non-equilibrium frontier growth. We find that the frontier of the growing cluster sketched out by the conserved charge appears {\it not} to fall in the same universality class as a simple random walk and iso-height lines of Kardar-Parisi Zhang. Instead, the frontier  fractal dimension is consistent with $5/4$, the value for the loop-erased random walk. Based on numerics and heuristic arguments we conjecture that the scaling limit of the frontier of the conserved charge cluster is described by an SLE($2$) conformally invariant curve, as opposed to SLE($8/3$) (self-avoiding walk) for the conventional cases.

\section{Stochastic dynamical charge hopping model}
\begin{figure}
    \centering
     \includegraphics[scale = 0.17]{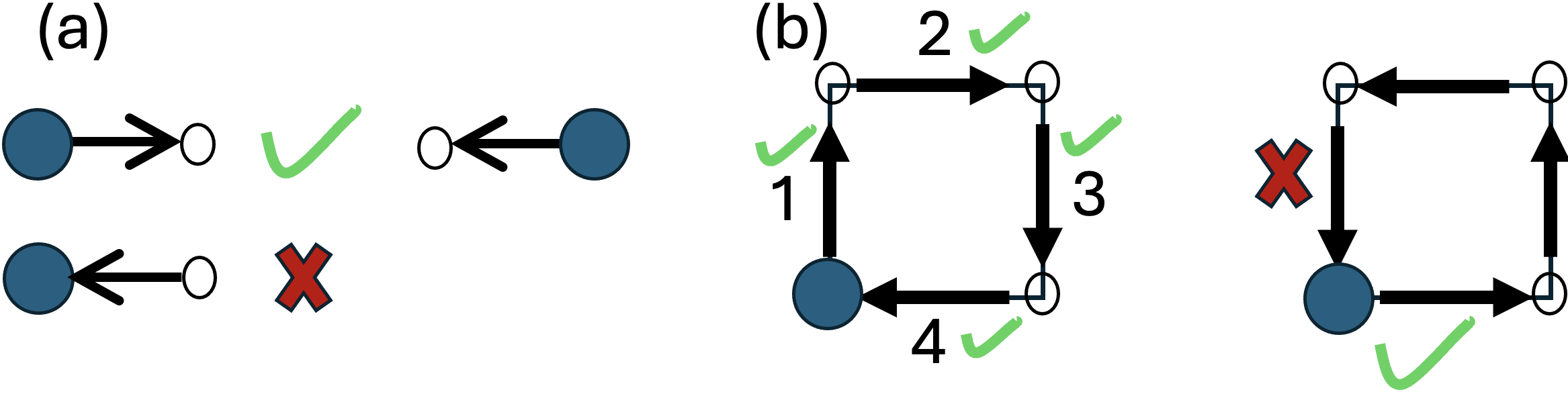}
    \caption{(a) Constraint on defect (large blue circle) motion imposed by the  Ising degrees of freedom (black arrows) on the links. A hop is allowed only if arrow points away from defect and neighboring site is empty. (b) Corresponding loop constraint: after a clockwise sequence of hops 1,2,3 and 4 the defect returns to its initial position, at which point the link variable flipped by its first hop blocks further clockwise hops.}
    \label{Fig1}
\end{figure}
We start with a random classical configuration with Ising degrees of freedom on links of a square lattice subject to a hard-core constraint at every vertex. Such constrained models can be represented as a gauge theory where every vertex satisfies a Gauss' law constraint. One can represent these Ising variables as arrows along the links, and the corresponding vertex constraint refers to the difference between the number of incoming and outgoing arrows. Versions of this constraint include the ice rule of two arrows pointing towards (in) and two away (out) from every vertex, and the dimer constraint of only one dimer covering every site, represented as a 3 in-1 our or 3 out-1 in for the respective sublattices of a bipartite (two inequivalent sets of sites) lattice.

We consider the  constrained classical hopping problem of a dynamical charge (defect). These are created by flipping an arrow chosen at random, resulting in two monopole defects for the ice problem and two monomer defects for the dimer problem. For clarity, we focus on the square ice problem for the rest of this paper but the general theory applies equally to the dimer or any similarly constrained model (see the discussion on \textit{implications for similar models} for quantitative considerations).  We fix one such monopole (defines the origin) and allow the other one to hop stochastically satisfying the condition presented in fig. \ref{Fig1}(a).  While appearing simple, such a hopping corresponds to a many-body dynamical problem due to the hard-core constraints, which can be expressed in the language of an emergent gauge theory. Although the underlying degrees of freedom are Ising-like (take on one of two values), the emergent gauge symmetry is U(1). This constraint is best understood pictorially. As shown in Fig. \ref{Fig1}(a), Gauss' law imposes the constraint that the dynamical charge can only hop across a link, if the arrow points along the direction of the hop. Besides its theoretical simplicity, such constrained hopping is relevant for the motion of defects (matter fields) in platforms such as artificial spin-ice (quantum simulators realizing lattice gauge theories) \cite{wang2006artificial,Articereview}. Due to the Ising links encoding a U(1) symmetry, such a hopping mimics that in the spin-1/2 U(1) quantum link model \cite{chandrasekharan1997quantum}, where the constrained hopping term would be the 3-body gauge-matter coupling.
 
For our classical analysis, in this work we do not consider coherent 4-spin plaquette flips. Such plaquette-flips, in the absence of any dynamical charge, lead to interference-induced disorder-free localization with unique spectral signatures \cite{DFL1,DFL2}. In the near future, the 3-body terms are  target for quantum simulators aiming to realize U(1) gauge theories. Our analysis of such incoherent stochastic dynamics clearly separates the role of hard-core constraints from that of quantum coherence. 

Dynamics of such frustrated models have been studied in a different context --- purely relaxational (model A \cite{hohhalpdyn}) and local dynamics of stochastic plaquette flips \cite{henley1997relaxation}. Such stochastic dynamics always results, in the long-time limit, to standard diffusion, in which the equilibration time for the plaquette flips has a dynamical exponent $z=2$. However, we show that the non-local nature of the fractionalized defects allows for a tunable dynamical exponent with asymptotic subdiffusion. The tunability can be exploited in novel synthetic platforms which can tune lattice-level constraints.

\section{Tunable long-time subdiffusion}
First, we present numerical data for the microscopic dynamics of the monopole hopping in the ice background. As claimed, we find that the asymptotic dynamics (long-time, long-distance) of the monopole is subdiffusive. As seen in Fig. \ref{Fig2}a (see parameter and other details of all numerical data for all figures, in appendix~\ref{numdetailsapp}), the mean-squared displacement of the dynamical charge has a sub-linear scaling with time, $\langle r^2 \rangle \sim t^{\beta}$, where $\beta = 0.86$, where $t$ is in units of number of attempted hops in the monte carlo and $\langle r^2 \rangle$ is in units of the lattice spacing. 

Such robust sub-diffusion on a simple Euclidean lattice, in the absence of disorder and away from any fine-tuned critical point is, as we will highlight in this paper, a unique feature of such many-body models with hard-core constraints.  

\begin{figure}
    \centering
    \includegraphics[scale = 0.11]{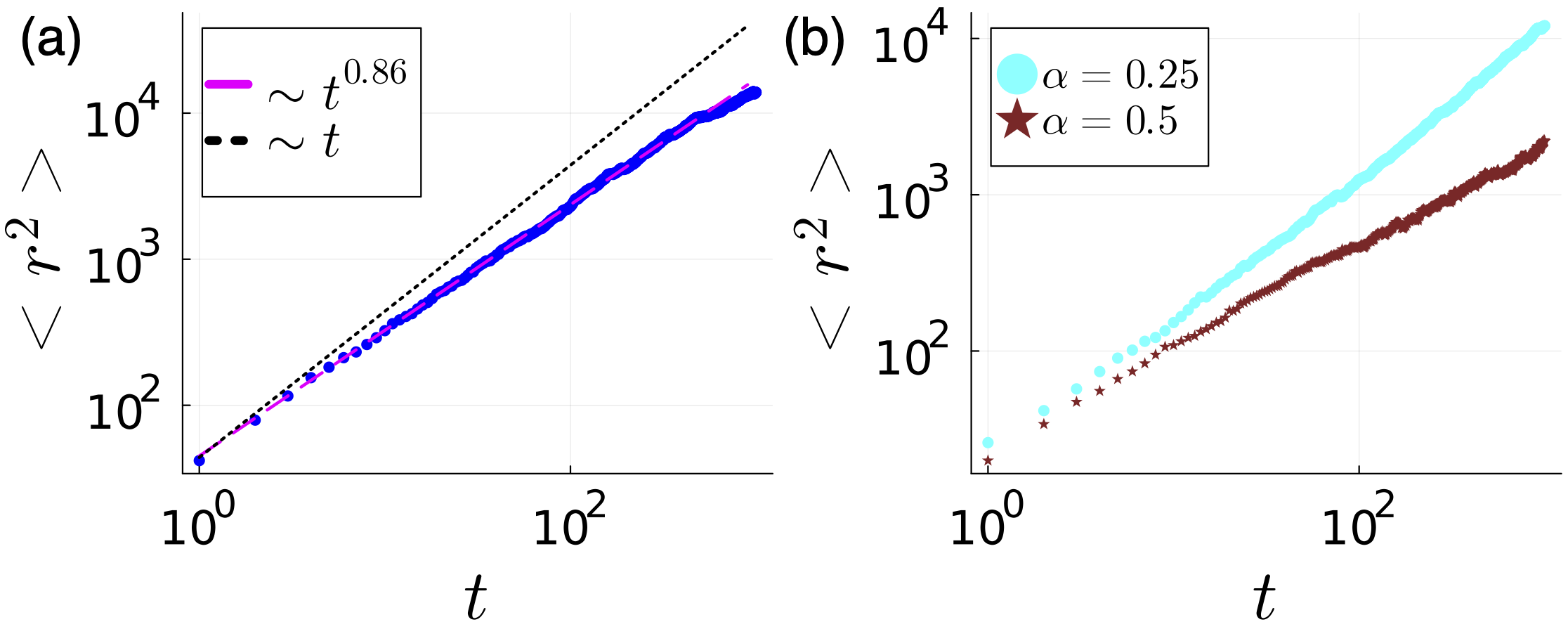}
    \caption{ a) Subdiffusion of the defect (monopole) in square ice with $\langle r^2 \rangle \sim t^{0.86}$ (dashed pink line is the numerically obtained best fit to the data in blue), for a $400 \times 400$ lattice, where $\langle r^2 \rangle$ is in units of lattice spacing and $t$ in units of attempted hops (monte-carlo time). b) Introducing a distribution of static charges (violations of the ice-rule), parametrized by $\alpha$, changes ground state correlations and also the long-time subdiffusion exponent. See end matter for numerical details.}
    \label{Fig2}
\end{figure}

Further, Fig.~\ref{Fig2}b, varying lattice-level constraints by adding static charges can continuously change the long-time subdiffusive exponent (see end matter for details on parametrization $\alpha$). Our results establish such continuously varying \textit{dynamical} exponents as another unique feature of such two dimensional constrained models (emergent gauge theories).
In the following sections, we provide detailed theoretical explanations from varied perspectives for the numerically observed anomalous dynamics of the microscopic defect hop problem. While the numerics is for the exact microscopic dynamics, we will show that one can develop two equivalent descriptions of such anomalous diffusion through i) an effective continuous time random walk framework and ii) a Langevin equation derived from the height field theory of such frustrated magnets.

\section{Dynamical caging and fractional diffusion}
First, we present an analytical, microscopic and partly phenomenological description of the asymptotic subdiffusive dynamics of the defect using the formulation of an effective continuous-time random walk (CTRW) \cite{montroll1965random,metzler2014anomalous}. Such a lattice-level description with slight modifications is applicable to all fractionalized defects in such frustrated models. 


\begin{figure}[t]
  \centering
  \includegraphics[width=0.3
  \textwidth]{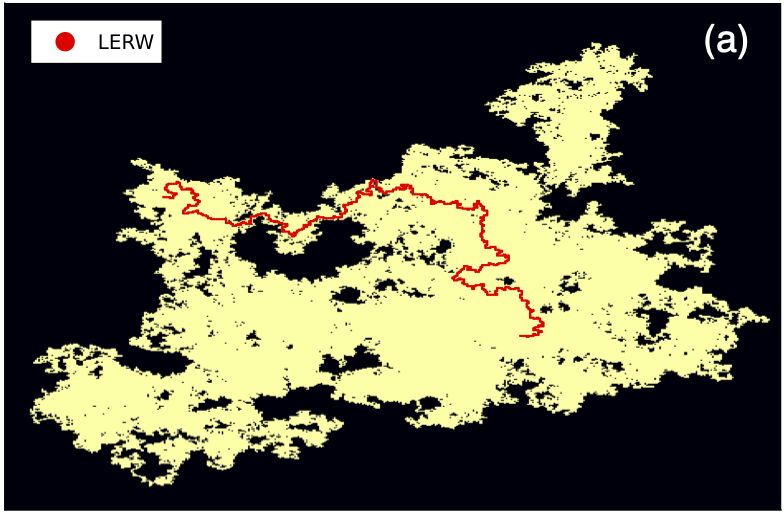}\\[1ex]
  \begin{minipage}[t]{0.2\textwidth}
   \vspace{0pt}%
    \centering
    \includegraphics[width=\textwidth]{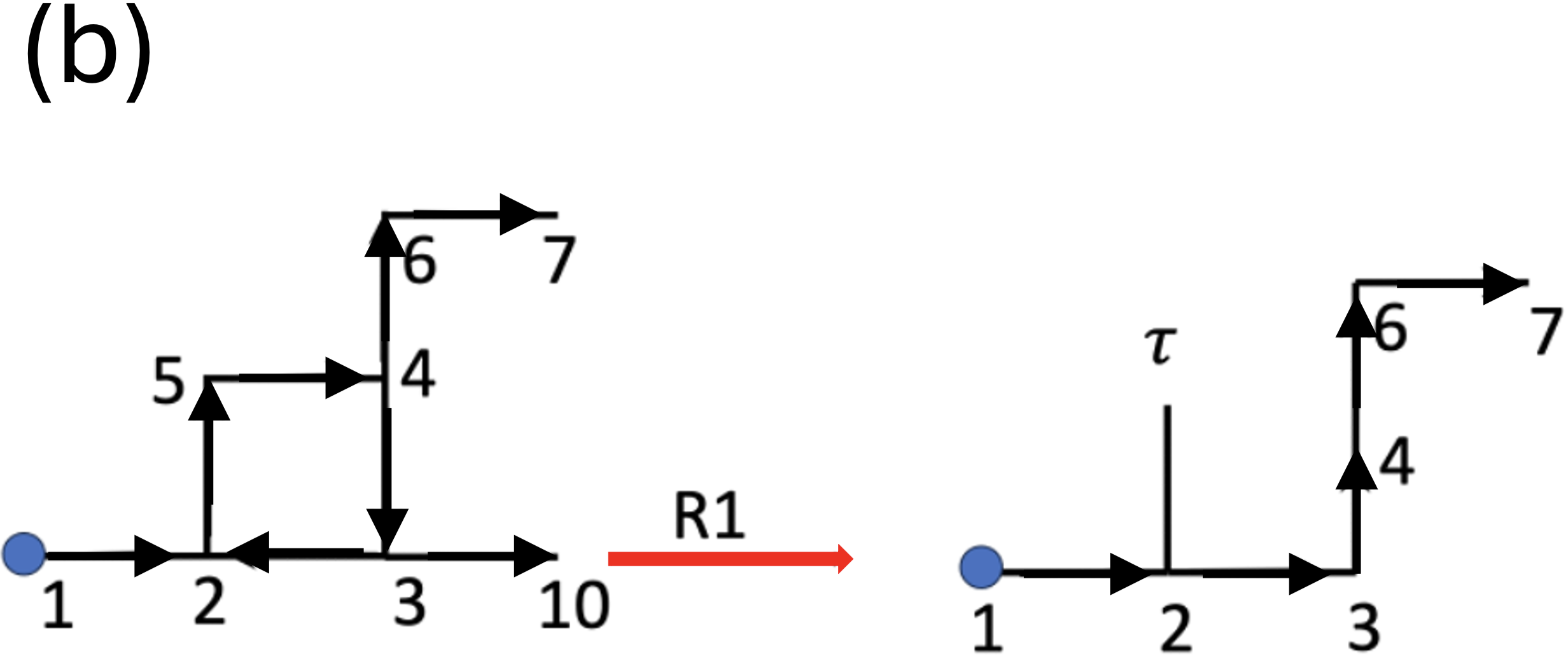}
  \end{minipage}
  \hfill
  \begin{minipage}[t]{0.23\textwidth}
   \vspace{0pt}%
    \centering
    \includegraphics[width=\textwidth]{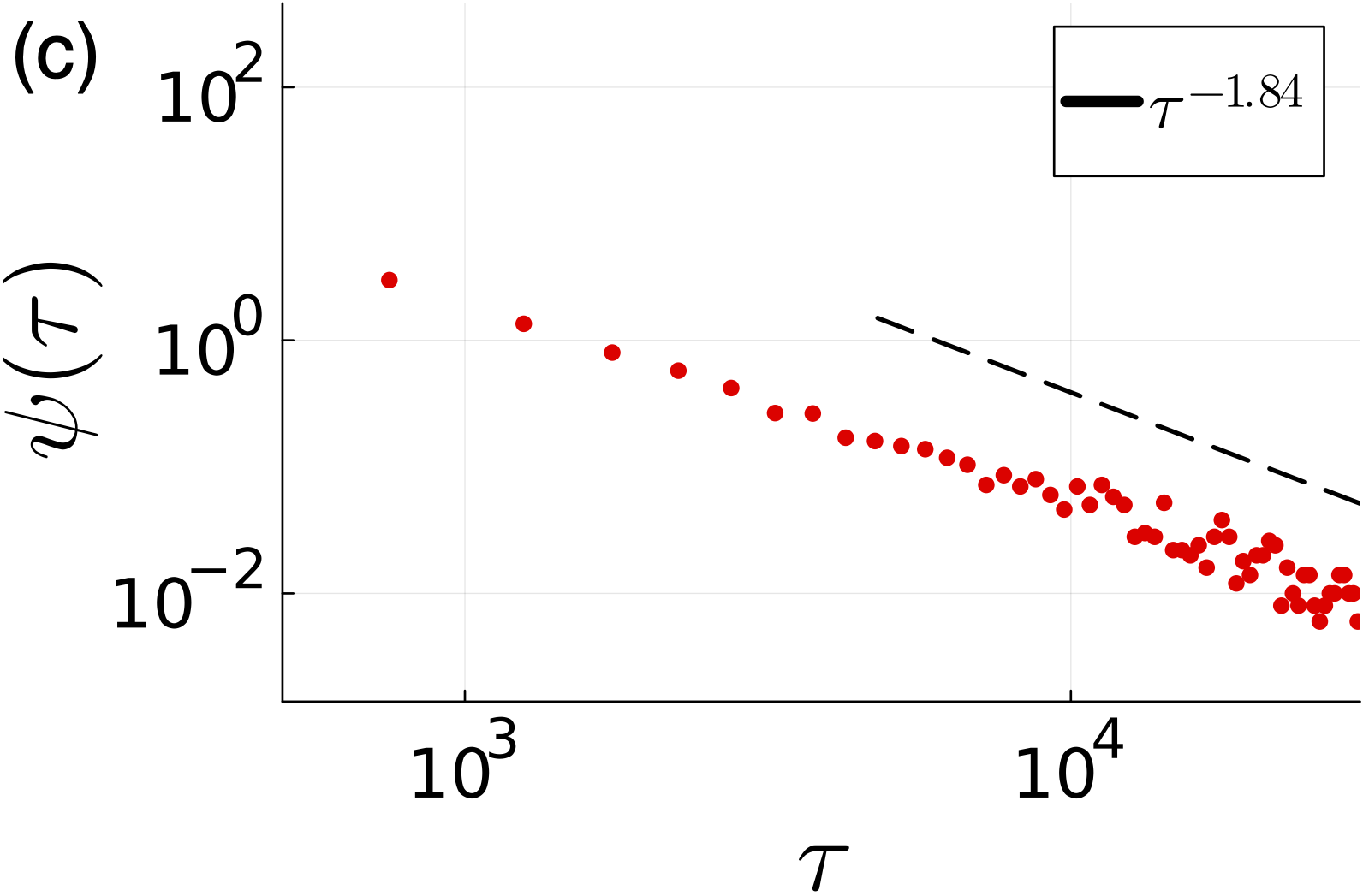}
  \end{minipage}
  \caption{a) The trace of the defect's walk (yellow) and the corresponding loop-erased random walk (LERW) (red) b) Effective description of a random walk $R1 \equiv (1,2,5,4,3,2,3,4,6,7)$ (all these hops are allowed, see Fig. \ref{Fig1})  as a loop-erased random walk with a waiting time $\tau$ c) Waiting time (erased loop-length) distribution, $\psi(\tau) \sim \tau^{-1.84}$, for the CTRW corresponding to the LERW of the dynamical charge for a $1000 \times 1000$ lattice.}
  \label{fig3}
\end{figure}

 A CTRW is governed by a random time scale, the waiting time between two jumps $\tau$; and a random length, the distance of the jump $s$. We consider the CTRW process in which these two random quantities are uncorrelated. We denote the corresponding probability density functions as $\psi (\tau)$ and $\phi(s)$. 
In situations where either $\langle \tau \rangle$ or $\langle s^2 \rangle$ diverges, anomalous diffusion may occur \cite{metzler2014anomalous}.

As demonstrated earlier, the Gauss' law constraint imposes a certain constraint of the random walk by not allowing the dynamical charge to circle around loops (of all length scales) along a single direction. As a result, any $n$-step random walk $r_w \equiv [r_0,r_1,\cdots, r_n]$ can be modelled as an $m$-step random walk $r'_w \equiv [r_0, (m-2) \ \rm{pos},r_n]$ with random waiting times $[\tau_1,\cdots \tau_{m-1}]$ (lengths of erased loops) as shown in fig. \ref{fig3}(b) (see end matter for details). The reduced path $r'_w$ is the loop-erased random walk (LERW) of $r_w$ which is obtained by removing all loops from the walk as soon as each one is formed \cite{lawler1999loop,schramm2000scaling}.


In fig. \ref{fig3}(c) we plot the distribution of the waiting times on the sites of the LERW for the dynamical charge (monopole) in the square-ice background. We see that it has a power law distribution whose asymptotic behaviour is given by $\tau^{-z}$ where $z \approx 1.84$. Hence, we can write 
\begin{equation}
    \psi(\tau) = \tau_0^{\beta}/\tau^{1+\beta}\,,
    \label{waitdist}
\end{equation}
where $\tau_0$ is some constant which is related to the generalized diffusion coefficient and $\beta \approx 0.84$. Importantly since $0 < \beta < 1$, Eq. (\ref{waitdist}) implies that the mean waiting time $\langle\tau\rangle$ diverges. Note, importantly, there is no intrinsic waiting time in this problem, the statement on diverging mean waiting time implies that the monopole dynamics is characterized by anomalously long retractions onto the sites of the LERW (see fig. \ref{fig3}(a)). This is what we refer to as dynamical caging. Such an effective caging mechanism and the anomalous dynamics is purely due to the  correlations within the ice-manifold of square-ice, as described below \cite{Liebice}. 

Further, one can show that a CTRW with waiting time distribution given in Eq.(\ref{waitdist}) can be represented by a modified Langevin equation which can be written (in the diffusion limit) as a fractional diffusion equation \cite{metzler2014anomalous} (also see appendix \ref{CTRWapp} )
\begin{equation}
    \frac{\partial}{\partial t}P(x,t) = K_{\alpha0} D_t^{1-\beta} \frac{\partial^2}{\partial x^2}P(x,t)
\end{equation}
where $K_{\alpha}$ is the generalized diffusion coefficient, $D_t$ is the Riemann-Liouville fractional operator acting on the spatial derivative of $P(x,t)$
and $x$ is a position on the loop-erased walk. From the above equation we get $\langle r^2 \rangle \sim t^{\beta}$ which agrees well with our numerics in fig. \ref{Fig2}(a). 

\section{Continuum description: Height mapping and diffusion in logarithmic potential}
The microscopic picture provided a semi-analytical description (requiring numerics on erased-loop size/waiting time distribution) of monopole subdiffusion,  we next provide a completely analytical coarse-grained description of this  dynamics.

This involoves a so-called height model, obtained via a mapping (familiar from  frustrated magnetism) to a  interface represented by a scalar height function $h(\bm{r})$. Crucially, this fluctuates due to residual entropy of the ice or dimer manifold of states. 
The resulting action for this coarse-grained field theory is
\begin{equation}
    S = \int \frac{K}{2} |\nabla h(\bm{r})|^2 + \cdots
    \label{heighteq}
\end{equation}
where $K$ is the stiffness of the height field and the additional terms are irrelevant (in the RG sense) for the critical (rough) phase corresponding to the critical ground state correlations in the ice-manifold (or dimer model) \cite{chalkernotes}. The presence of a monopole, corresponds to vortex-antivortex pairs (screw dislocations) in the height model.
More importantly, in two dimensions, the interaction between a vortex and anti-vortex is then given by an attractive potential V(r) = $K \rm{ln}(r/a)/\pi$ \cite{chalkernotes}, where the coefficient of the regularized logarithmic potential arises from a combination of the height stiffness and the burgers vector of the defect (defect charge). This potential is purely entropic in nature.

Such an analysis makes the difference between plaquette flips and the presence of a monopole clear: while plaquette flips affect the height of only neighbouring plaquettes, a monopole (fractionalized defect) has a non-local effect through the logarithmic potential. Moreover, the macroscopic dynamics of our defect can be expressed as a Langevin equation with a logarithmic potential. Exact solutions of the respective Fokker-Planck equations have been obtained using tools such as infinite covariant densities \cite{Kesslerdiff,dechant2011solution}. The mean-squared displacement scales as $\langle r^2 \rangle \sim t^{1.5-K/\pi}$. From exact results of the ice model ground state correlations on the square lattice, we know that, $K = 2\pi/3$ and hence $\langle r^2 \rangle \sim t^{0.83}$, in good agreement with numerics in Figs. \ref{Fig2}(a), \ref{fig3}(c) and the analysis in the previous section. A similar exponent has been obtained from a scaling ansatz and a phenomenological consideration of the different types of noise in the ice-manifold \cite{nisoli2021color}.

Moreover, diffusion in a logarithmic potential is an extremely rich problem, since as noted initially by Bray \cite{Braylog}, the noise term in the Langevin equation is \textit{exactly marginal} in the renormalization group sense. This leads to tunable (discussed in next section) dynamical exponents for subdiffusion in the long-time limit as we see in Fig. \ref{Fig2}(b). Most importantly, the dynamical exponent depends on the coefficient of the logarithmic potential, which in turn depends on static correlations of the ground state. If instead of a logarithmic potential one had a power-law potential, as in three-dimensional spin ice, then in the asymptotic limit one would recover standard diffusion (in the absence of any fine-tuning).

\section{Implications for higher-spin and similar models: tunability and UV-IR mixing}
Higher spin versions of these link models in two dimensional settings have been considered in the context of the antiferromagnetic Ising model on the triangular lattice, which has a dimer representation on the honeycomb lattice \cite{Zheng97}. In the corresponding height-field theory, the stiffness has been shown to increase on increasing spin \cite{Zheng97}, implying a change in the coefficient of the logarithmic potential and hence, according to our analysis, a different subdiffusion exponent. Alternatively, adding  interaction terms can also change the ground state correlations and thereby the subdiffusion exponent. Our results therefore identify such frustrated systems as possibly the simplest examples of tunable subdiffusion. Moreover, since as we see in Fig. \ref{Fig2}(b), changing lattice-level constraints affects asymptotic dynamics. Hence, such systems also provide a platform for probing UV-IR mixing in two dimensions.

\section{Frontier growth conjecture and SLE.---}
We change perspectives and view the fractionalized defect hopping as a non-equilibrium growth process. The relevant growth is that of the \textit{frontier} of the cluster sketched out by the monomer, defined as the boundary of the cluster with all the holes filled in. We analyze such growth through the lens of Schramm-Loewner evolution (SLE) \cite{cardy2005sle}.

Due to their characterization of simple curves via a single parameter, $\kappa$, SLEs lend themselves naturally to describe growth in two dimensions \cite{bauer20062d}. Discrete interface growth models such the ballistic deposition, restricted solid-on-solid and Eden models are all believed to fall into the two-dimensional Kardar-Parisi-Zhang (KPZ) \cite{KPZ} universality class with iso-height lines in these models converging (in the scaling limit) to SLE($8/3$). Moreover, the frontier of regular Brownian motion, as conjectured by Mandelbrot and proved later by Werner, Lawler and Schramm using SLE methods \cite{Lawlerproof}, also converges to SLE($8/3$) --- the universality class of the self-avoiding walk. 

\begin{figure}[t]
  \centering
  \includegraphics[width=0.3
  \textwidth]{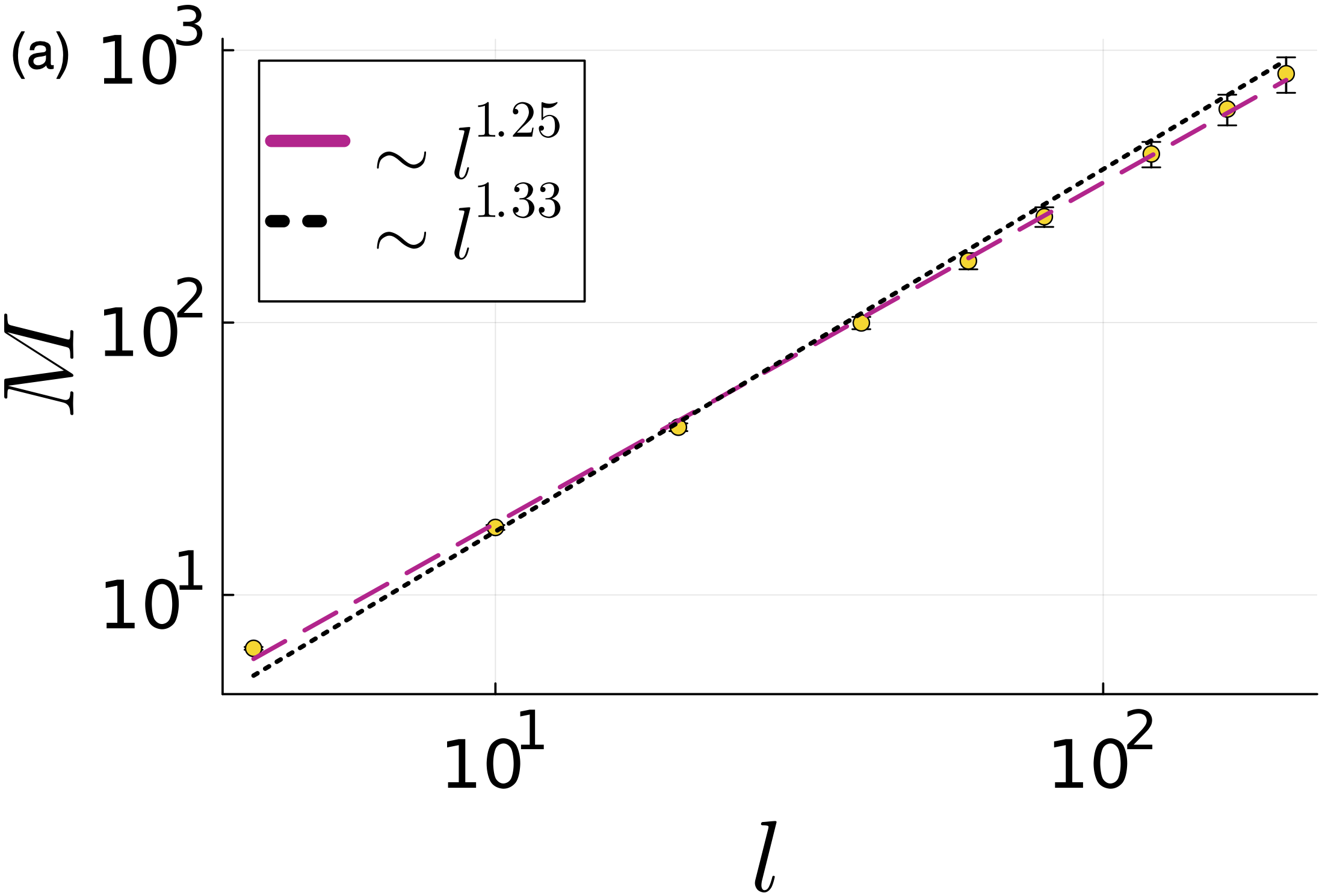}\\[1ex]
  \begin{minipage}[b]{0.4\textwidth}
    \centering
    \includegraphics[width=\textwidth]{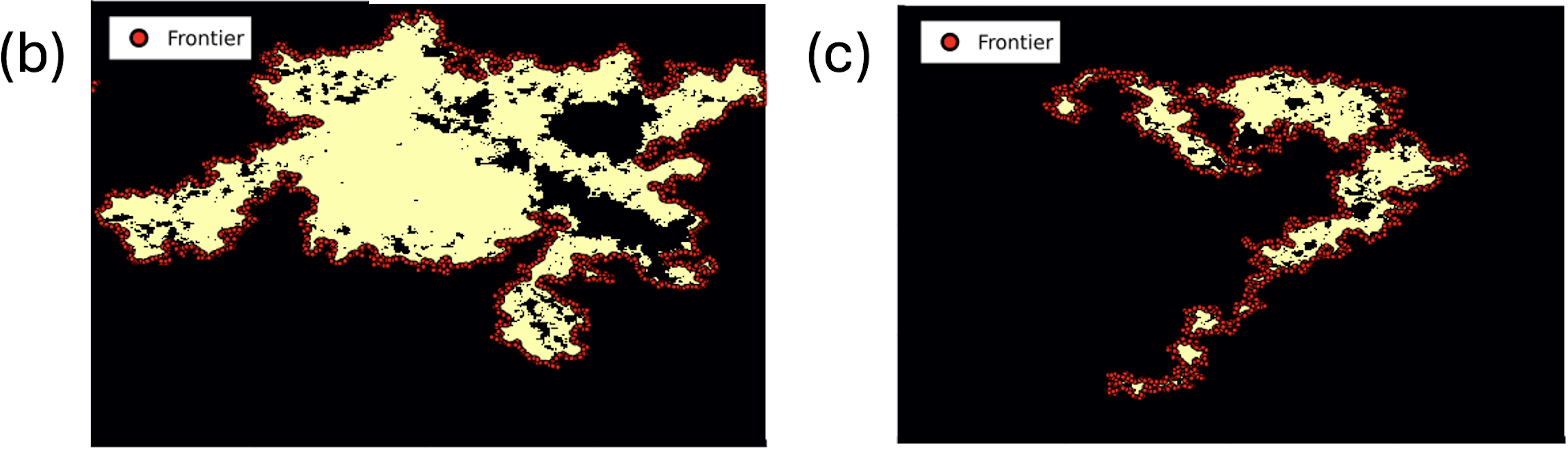}
  \end{minipage}
  \caption{(a)Fractal dimension of conserved charge motion frontier (in a $600 \times 600$ lattice) is "thinner" than that of standard Brownian motion, $d_f = 1.25 < 4/3$, just as the LERW is "thinner" than the SAW \cite{lawler1999loop}. (b) Typical cluster of the conserved charge is more "space-filling" than that of a simple random walk (c), as expected from duality of SLEs.}
  \label{Fig4}
\end{figure}
  
 First, we obtain numerical data for the fractal dimension of the frontier in Fig. \ref{Fig4}a (see end matter for details). The best linear fit to this data gives a fractal dimension $d_f \approx 1.25$ in  agreement with the rigorously known growth exponent of the loop-erased random walk, $5/4$ \cite{majulerw,kenyon2000asymptotic}.
 Such a fractal dimension is also consistent with an SLE($2$) description, since $d_f = 1 + \kappa/8$ for $\kappa <= 8$.
 
Second, purely from visual inspection of Figs. \ref{Fig4}b-c, one can infer that the dynamical charge random walk is more "space-filling" than that of a standard random walk. In SLE, this property corresponds to a higher diffusivity $\kappa$. In SLEs, there is a conjectured duality, first suggested by Duplantier, stating that frontier of the hull (cluster with all internal holes filled in) which scales as SLE($\kappa'$) is described by SLE($16/\kappa'$) \cite{Duplantierduality}. If the frontier has a lower diffusivity, then the corresponding hull is more space-filling, hence this simple observation already implies that the frontier growth is described by some $\kappa < 8/3$.

Besides, numerical observations we present a heuristic argument for the deviation from SLE($8/3$) behaviour. Say that the dynamical charge starting at the origin, after time $t$ is at the some point $r_t$. By definition, this point belongs to the frontier of the cluster at time $t$. Now, using the arguments presented in earlier sections, one can consider the growth of the cluster effectively as the growth of the loop-erased walk. Unlike standard Brownian motion, the self-avoided walk or random percolation clusters, the loop-erased random walk is non-local, i.e heuristically, the growth of the walk "feels" the effect of the erased loops at all length scales. Hence the growth of the frontier in our problem is described by a new universality class, different from standard 2d KPZ and SLE(8/3).

\section{Finite density of defects}
\begin{figure}
    \centering
    \includegraphics[scale = 0.11]{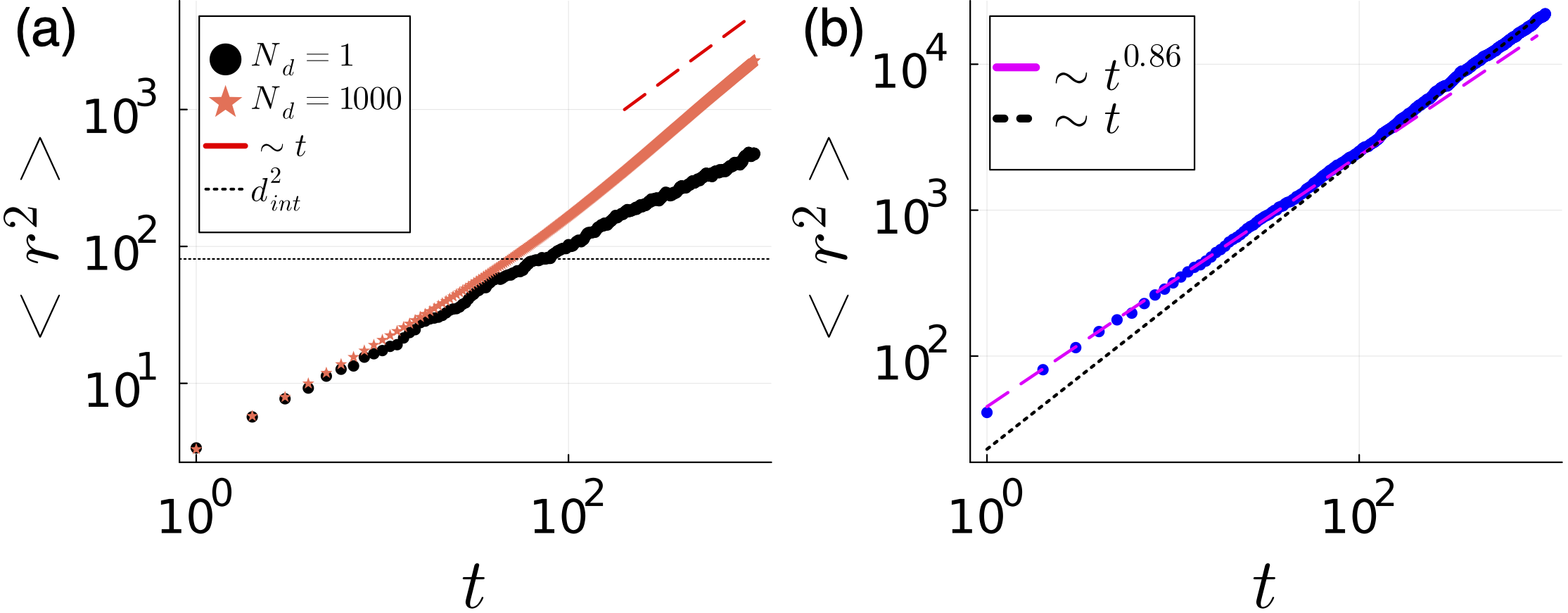}
    \caption{Average mean square displacement for a finite density of defects ($N_d$ = number of mobile defects), displaying subdiffusion upto average inter-particle spacing ($\sim$Debye screening length). (a) For an initial configuration drawn from the ensemble generated by $\alpha = 0.5$ on a $400 \times 400$ lattice (see Fig. \ref{Fig2} (b) and numerical details in end matter). (b)  For an initial configuration drawn from the ice manifold with $N_d = 1000$ on a $1000 \times 1000$ lattice. Panel (a) visualizes the crossover to diffusion in a clearer fashion due to the greater difference in long-time exponents, as compared to that in square ice.}
    \label{fig5}
\end{figure}
We now consider the dynamics of a finite density of defects. Our coarse-grained macroscopic description would propose that  as long as the defects feel the entropic logarithmic potential, the subdiffusion  persists. This  occurs  until the mean square displacement is of the order of the average inter-particle spacing. Beyond such a regime, an analogue of Debye screening~\cite{krauthmoessner} would restore diffusion, as then the problem would reduce to standard Brownian motion. 

As we see in Fig.~\ref{fig5} for both the monopole in square ice case (\ref{fig5}b) and for the defect in the presence of static charges (\ref{fig5}a), the average mean-squared displacement per particle displays subdiffusive behaviour up to a distance (and the corresponding timescale) of the order of the average interparticle spacing. Beyond this, there is a smooth crossover to standard Brownian motion.

\section{Discussion and Outlook} We have shown that a dynamical charge subject to a gauge constraint in two dimensions, arising naturally in discrete link models, can lead to anomalous dynamics and non-equilibrium growth. Such dynamics arises due to critical correlations of the ground state manifold in such models, which induce anomalously long retractions and a form of dynamical caging. Besides rich theoretical phenomenology, such systems are being realized in various cutting-edge quantum simulation as well as artificial spin-ice systems. We highglight this more below in the end matter. Quantum extensions of the above problem have interesting consequences for non-ergodicity as the Hilbert space configuration can be well approximated by Galton-Watson trees, i.e, a tree with disorder in coordination number. Such a Hilbert space structure could be amenable to realizing nonergodicity in infinite volume. Further, three-dimensional extensions of the above work, would be interesting theoretically as well as experimentally from the point of view of frustrated magnets such a pyrochlore spin-ice. 

\begin{acknowledgments} {\it Acknowledgements.---} We thank Abhinav Sharma, Benoit Doucot, Christoph Weber, Claudio Castelnovo, Hongzheng Zhao, Shubhro Bhattacharjee and Sthitadhi Roy for valuable discussions. This work was supported by the Deutsche Forschungsgemeinschaft via Research Unit FOR 5522 (project-id 499180199), as well as the cluster of excellence
ct.qmat (EXC-2147, project-id 390858490).
\end{acknowledgments}

\newpage

\appendix
\section{Experimental platforms}
Remarkable recent progress has been made in realizing the PXP model in two dimensional platforms using the Rydberg blockade mechanism. To observe such diffusion we propose an implementation in which the Rydberg atoms are placed on the links of a honeycomb lattice. The Rydberg blockade mechanism is then implemented by choosing a blockade radius of $\sqrt{3}a/2$ ($a$ - lattice constant), which generates dimer-type constraints, as has recently been achieved \cite{semeghini2021probing}. Another promising platform for the observation of such emergent phenomena is artificial spin ice - arrays of nano-magnetic islands in two dimensions \cite{moller2006artificial,perrin2016extensive}. Moreover, recent experiments have also demonstrated the ability to do real-time imaging of defect motion~\cite{farhan2019emergent}.

Finally, there is a whole host of other platforms, such as optical lattices in ultracold atoms, superconducting circuits etc. which can realize lattice gauge theories. There has been impressive recent progress in such pursuits with a large push to realize two-dimensional lattice gauge theories in the lab \cite{schweizer2019floquet,zhou2022thermalization,semeghini2021probing,meth2023simulating,martinez2016real,cochran2024visualizing,yang2020observation,gonzalez2024observation}. Our results would be a concrete test to see if such emergent gauge theories, and thereby the ensuing rich dynamical phenomena, are indeed realized in such settings.

\section{Continuous Time Random Walk formalism}
\label{CTRWapp}
We provide some more details of the continuous time random walk (CTRW) formalism that arises as an effective description for our fractionalized charge hopping problem resulting in an effective fractional diffusion equation in the asymptotic limit. 

This description emerges as follows: Due to the hopping constraint depicted in fig. \ref{Fig1}(b), every time the charge traverses a loop (of arbitrary size) it cannot continue in the same clockwise/anticlockwise direction. The available options are to exit the loop or to retrace along the path it took to complete the loop. Hence, a natural effective description of such dynamics is to erase every loop, as soon as it is formed in the trajectory of the dynamical charge, and introduce a waiting time $\tau$ at the point of erasure (the point where the loop is first formed).

The dynamics is then described as follows: the charge moves along the loop-erased path (formed by erasing every loop chronologically, i.e as soon as a loop is formed in the trajectory) but with a waiting time on every site of this path (time spent at this site before jumping to the next site) characterized by the length of the erased loop. Such a dynamics represents that of a CTRW. As mentioned in the main text, the CTRW is characterized by two quantities: the mean waiting time $\langle \tau \rangle$ and the step-size variance $\langle s^2 \rangle$. These quantities are expressed as:
\begin{equation}
\begin{split}
    \langle \tau \rangle = \int_0^{\infty} \tau \psi(\tau) d\tau \\
    \langle s^2 \rangle = \int_{\infty}^{\infty} s^2 \phi(s) ds
\end{split}
\end{equation}
where $\psi(\tau)$ and $\phi(s)$ are the waiting time and step-size distributions. In our constrained hopping problem, the step-size is fixed to the lattice spacing. The crucial step in our CTRW modelling is representing the erased-loop sizes as effective waiting times. \textit{There is no in-built waiting time in our problem}. However, due to the constraints mentioned in the main text and above, the erased loop distribution acts as an effective waiting time distribution, giving a waiting time to every site of the loop-erased walk and thereby allowing us to use CTRW methods to model the asymptotic subdiffusion.

As we saw in the main text, the waiting time (erased-loop size) distribution is obtained numerically, and is a heavy tailed distribution, i.e $\psi(\tau) \sim \tau^{-1.84}$. Such heavy tailed distributions of the form $\sim \tau^{-(1+\beta)}$, for $0<\beta < 1$ result in diverging mean-waiting times. Such a diverging mean waiting time, as mentioned in the main text for our hopping problem, represents a dynamics characterized by anomalously long retractions to points on the loop-erased path - dynamical caging.

Starting from such a heavy-tailed distribution for the waiting times, one can show that the mean-squared displacement is given by $\langle r^2 \rangle \sim t^\beta$. Further, one can obtain the continuum stochastic equation describing the asymptotic dynamics in the form of a fractional diffusion equation of the form.

\begin{equation}
\begin{split}
    \frac{\partial}{\partial t}P(x,t) = K_\alpha D_t^{1-\beta} \frac{\partial^2}{\partial x^2}P(x,t) \\
    D_t^{1-\beta} = \frac{1}{\Gamma (\beta)}\frac{\partial}{\partial t}\int_0^t \dfrac{P(x,t')}{(t-t')^{1-\beta}}dt'
\end{split}
\end{equation}
where $D$ is known as the Riemann-Liouville fractional operator. The above equation reduces to the standard diffusion equation in the limit $\beta \rightarrow 1$. Thus while the asymptotic dynamics of a simple random walk results in a continuum diffusion equation, the asymptotic dynamics of such fractionalized defects in frustrated models result in effective fractional diffusion equations.

\section{Height models and logarithmic potential}
Besides the CTRW, one can also use the height model to obtain a coarse-grained and fully analytical description of the asymptotic subdiffusion. Further, the height description elucidates the role of the ground-state correlations in the dynamics. 

The height model describes the residual entropy of a class of frustrated models in terms of a rough surface (see \cite{chalkernotes} for a very clear exposition). The height field lives on the dual lattice, which in our case is defined by the centres of the square plaquettes. The height difference between adjacent plaquettes is determined by the direction of the arrow on the link shared by them for the ice model, or the presence/absence of a dimer on the shared link for the dimer model. The prescription for the height differences are chosen such that the net height difference for any contractible closed loop on the dual lattice vanishes.


Such a  microscopic map identifying ice/dimer with a unique height configuration encodes the hardcore constraint of the ice/dimer model. This microscopic height field is then amenable to coarse graining  into $h(r)$, to describe the long-wavelength physics. 
This results in  Eq. \ref{heighteq}, a simple Gaussian field theory. Other terms are irrelevant (in an RG sense) in the rough phases of such height models which capture the residual entropy of dimer and ice models in the regime we are interested in. 

The fluctuations of this height field have a logarithmic divergence, i.e $\langle |h(0)-h(r)|^2\rangle \approx \text{ln}(r/a)/(\pi K)$ ($a$ is the lattice spacing) from which one can calculate the Ising correlators as well. The height description also incorporates the description of defects/excitations. A single spin flip defect creates two monopoles, which when separated by a distance $r$ are described as two screw dislocations of opposite vorticity in the height description. This highlights the effectively non-local nature of the defect. While a plaquette-flip excitation only changes the height locally, the monopole at the origin leads to a long-range disturbance following a $1/r$ Gauss' law. As a result, two monopoles when separated by a distance $r$, have an entropic attractive potential $V(r) =  V_0 \text{log} (r/a)$. The coefficient of this potential is determined by the Burgers vector of the screw dislocation ("magnetic charge" of the defect in the field theory language) and the stiffness of the height field. The latter quantity is directly related to the defect-defect correlations (monopole/monomer or others) in the rough phase. From known exact results in square ice, one finds that $V_0 = 2/3$, which as shown in the main text gives us $\langle r^2 \rangle \sim t^{0.83}$. Since the coefficient of the log-potential can be tuned continuously (e.g. by adding an interation between dimers),  the subdiffusion exponent is also straightforwardly continuously variable.

\section{Numerical details for figures:}
\label{numdetailsapp}
Fig. \ref{Fig2}(a): We start with a random state from the ice-manifold for a $400 \times 400$ square lattice. We then choose a link at random and flip it. This creates a "charge" at a 3 out-1 in vertex and another at a 3 in-1 out vertex. We fix the latter and let the former hop according to the rules laid out in the main text. The m.s.d in Fig. \ref{Fig2}(a) is obtained by averaging over 500 different initial states, where for each state we perform stochastic hops of the monopole. Each time step is defined as 100 attempted hops in the Monte-Carlo procedure and $\langle r^2 \rangle$ is in  units of the lattice spacing. We then plot the best-fit for the data shown in the main text which gives us an exponent of $0.86$. 
The nominal statistical error from the fit of this quantity is tiny (in the third decimal digit), so that the actual error, which we estimate to be around $0.03$, is dominated by systematic issues, such as possible corrections to the asymptotic behaviour. 

Fig. \ref{Fig2}(b) calculate $\langle r^2 \rangle$ in the same way, but instead of starting from a random configuration in the ice manifold, the initial state contains a density of defects violating the ice rule parametrised by $\alpha$.
On each link, we assign a probability $\sin^2{(\alpha+\pi/4)}$ of the link variable being the same as it would be if the arrows were arranged in one of the two clockwise/anticlockwise checkerboard configurations, and a probability $\cos^2{(\alpha+\pi/4)}$ of the link variable being the same as it would be if the arrows were arranged in the other possible one. Each value of $\alpha$ gives us an ensemble of configurations and we average over 500 different initial configurations for each $\alpha$ 

Fig. \ref{fig3}(c) analyses the resulting statistical properties of the trajectory of the mobile defect. The loop-erased path and the corresponding distribution of erased loops are obtained. The length distribution of the latter yields the effective waiting time, $\tau$, distribution, $\psi(\tau)$. Data is shown for a $1000 \times 1000$ square lattice, keeping track of the trajectory for $10000$ time units, i.e.\ $10^6$ attempted hops. For the plot, we bin the data in bins of length 500 (in units of $\tau$) and plot the average of all data corresponding to all the $\tau$s in each bins. We fit the range of data covered by the dashed curve to focus on the tail.

Fig. \ref{Fig4}(a): Again starting from a random-ice state, we  record the cluster sketched out by the monopole after $65000$ time-units ($65*10^5$ attempted hops) in a $600 \times 600$ lattice. The cluster is defined as the conencted network of all bonds of the lattice across which the monopole has hopped at least once. We identify the frontier of the cluster by recording the outermost boundary of the cluster, i.e the perimeter with all the holes filled in (see Fig. \ref{Fig4}(b)). We calculate the local fractal dimension of the frontier using the box-counting method, where we choose a point on the frontier at random as the center of the box and then calculate the number of frontier points, $M$, that lie within a square of length $l$. We choose $l = [4,10,20,40,60,80,120,160,200]$ and  obtain the fractal dimension using the relation $M \sim l^{d_f}$. Both the stochastic and systematic errors we estimate to be around 1-2\% in the last digit.

Fig. \ref{fig5}: In (a) we use the same parametrization as Fig. \ref{Fig2} (b) with a random state from the ensemble generated by $\alpha = 0.5$ on a $400 \times 400$ lattice, and measure $\langle r^2 \rangle$/$N_d$ after averaging over 500 initial states ($N_d$ = number of mobile defects). We choose $N_d$ random vertices which satisfy the ice-rule, flip a bond (chosen randomly) attached to the site, and then perform the same procedure as in Fig. \ref{Fig2}, for the constrained hopping. Here time is in units of $10N_d$ attempted hops. In $(b)$ we do the same as $(a)$ for $N_d = 1000$, but now starting from a random ice configuration on a $1000 \times 1000$ lattice as opposed to configurations in the ensemble defined by $\alpha = 0.5$ in (a).
\bibliography{Bib.bib}

\end{document}